\documentclass[12pt,a4wide]{article}
\usepackage{graphicx}
\usepackage{amsmath}
\usepackage{amssymb}
\usepackage{bm}
\begin{document}
 
 

\title{
\bf \large \bf Restricting profile function of hedgehog Skyrmion\\[30pt]
\author{Jun Yamashita$^{1}$ and Minoru Hirayama$^{2}$\\[10pt]
\\{\normalsize $^{1}$ {\sl Department of Physics, University of Toyama, Gofuku 3190,}}
\\{\normalsize {\sl Toyama 930-8555, Japan}}
\\{\normalsize {\sl Email : fxjun@jodo.sci.u-toyama.ac.jp}}\\
\\{\normalsize $^{2}$  {\sl Department of Physics, University of Toyama, Gofuku 3190,}}
\\{\normalsize {\sl Toyama 930-8555, Japan}}
\\{\normalsize{\sl Email : hirayama@sci.u-toyama.ac.jp}}\\[0.5cm]}}
\date{June 2006}

\maketitle

\begin{abstract}
The profile function for the hedgehog Skyrmion is investigated. After discussing how the form of the profile function is restricted by the field equation, the static energy is numerically calculated. It is found that the profile functions considered here sometimes give the static energy smaller than previous ones.
\end{abstract}

The Skyrme model\cite{Skyrme} is an attractive candidate for the effective theory of Quantumchromodynamics (QCD)\cite{Witten, Jackson}. Its static version is defined by the static energy functional 
\begin{align}
E=&\frac{1}{24\pi^2}\int d^3x \left[- \mbox{tr}(R_iR_i)-\frac{1}{8}\mbox{tr}\left([R_i,R_j] [R_i,R_j]\right)\right],
\\
R_i=&(\partial_iU)U^{\dagger},
\end{align}
where $U=U(\bm{x})$ is an element of $SU$(2). We are here adopting the normalization of $E$ used in refs.\cite{Battye1, Battye2, Manton}.
The important topological quantity associated  with the Skyrme model is the baryon number 
\begin{align}
B=-\frac{\epsilon_{ijk}}{24\pi^2}\int d^3x~\mbox{tr}(R_iR_jR_k).
\end{align}
With these definitions, we have the Faddeev-Bogomolny inequality $E\geq\left|B\right|$.\\

We here restrict ourselves to the case of $B=1$ and assume that $U(\bm{x})$ obeys the hedgehog Ansatz:
\begin{align} 
U(\bm{x})= \exp\left[if(r)\hat{\bm{x}}\cdot\bm{\tau}\right],\quad r=|\bm{x}|, 
\end{align}
where $\bm{\tau}$ denotes the triplet of Pauli matrices. The true profile function $f(r)$ minimizes $E$. It is known that $E$ can be written as\cite{Manton}
\begin{align} 
E=&\frac{1}{3\pi}\int_0^{\infty}\varepsilon(r)dr, \label{eqn:StaticEnergy}\\
\varepsilon(r)=&r^2f'(r)^2+2\sin^2f(r)[f'(r)^2+1]+\frac{\sin^4f(r)}{r^2},
\end{align}
where $f^{\prime}(r)$ denotes the derivative of $f(r)$ with respect to $r$. The field equation for $f(r)$ is given by 
\begin{align} 
\left[r^2+2\mbox{sin}^2f(r)\right]f''(r)+2rf'(r)+\sin 2f(r) \left[f'(r)^2-1-\frac{\sin^2 f(r)}{r^2}\right]=0.
\label{eqn:FieldEq}
\end{align}

The static configurations which minimize $E$ for a given $B$ is called Skyrmions. In the case of $B$ =1, Battye and Sutcliffe\cite{Battye1} began their numerical analysis with the initial trial configuration 
\begin{align} 
f(r)=4\mbox{Arctan}(\exp[-r])
\end{align}
which gives $E$=1.24035. It was concluded that the energy of the $B=1$ Skyrmion is equal to 1.232\cite{Battye1}. In the later literatures, we find $1.2322$ \cite{Battye2, Manton}. The value $1.2322$ is sometimes referred to as "true" \cite{Houghton, Piette}. Krusch\cite{Krusch} investigated Skyrmions on the 3-sphere $S^3$ with the radius $L$. In the flat space limit ($L\to\infty$ limit), he obtained the profile function 
\begin{align} 
f(r)=\pi-2 \mbox{Arctan}(kr)
\end{align}
where $k$ is an adjustable constant.\\

In this paper, we first discuss how the functional form of the profile function is restricted by the field equation. It is then important to know how the adjustable parameters come in the profile function. We investigate a method to introduce the two parameters in the profile function. We then calculate numerically the minimal value of the static energy under the variation of the two parameters. Though numerical calculations aided by Mathematica package, we obtain the static energy of hedgehog Skyrmion comparable with or smaller than the above values.\\

To see  the mathematical properties of $f(r)$ implied by the field equation involving a transcendental function of $f(r)$, it is convenient to transform it into an algebraic differential equation. It is also convenient to define an independent variable which varies in a finite interval as $r$ ranges from 0 to $\infty$. Thus we introduce $v(z)$ and $z$ by
\begin{align}
v(z)=& \mbox{tan}^2f(r), \label{eqn:Vz}\\
z=&\frac{r^2}{r^2+2}. 
\end{align}
Then we have the field equation
\begin{align}
\frac{d^2 v}{dz^2}&-\frac{1}{2}\left[\frac{3}{v-1}+\frac{1}{v}-\frac{1}{v-z}\right]\left(\frac{dv}{dz}\right)^2 \nonumber\\
&+\frac{1}{2}\left[\frac{1}{z-1}+\frac{1}{z}-\frac{2}{v-z}\right]\frac{dv}{dz}+\frac{v\left[v(z+1)-2z\right]}{2z^2\left(z-1\right)^2\left(z-v\right)}=0
\label{eqn:FieldEqz}
\end{align} 
and the static energy
\begin{align}
E&=\frac{1}{3\sqrt{2}\pi}\int_0^1\left[\frac{\left(z-v\right)\sqrt{z\left(1-z\right)}}{v\left(v-1\right)^3}\left(\frac{dv}{dz}\right)^2+\frac{v\left(3zv-4z+v\right)}{2\left(v-1\right)^2\sqrt{z^3\left(1-z\right)^3}}\right]dz.
\end{align} 
 
The field equation Eq.(\ref{eqn:FieldEqz}) allows us to assume that, in the neighborhood of $z_0$, there is a solution of the form
\begin{align}
v(z)=& \sum^{\infty}_{j=0}v_j (z-z_0)^{\alpha+j},\quad v_j:\mbox{const.} (j=0,1,2,\cdots)
\end{align}  
Then the procedure known as the leading order analysis in the theory of nonlinear differential equation is applicable \cite{ARS}. If we assume $\alpha<0$ and $z_0\neq 0, 1$ and suitable the above expression for $v(z)$, the l.h.s. of Eq.(\ref{eqn:FieldEqz}) becomes $-\alpha\left(\alpha+2\right)v_0\left(z-z_0\right)^{\alpha-2}/2+\cdots$, where $\cdots$ represents terms less singular at $z=z_0$. We see that $\alpha$ must be $-2$. Though similar considerations in the cases $z_0=0$ and $z_0=1$, we conclude
\begin{align}\begin{cases}
&\alpha= 1 \quad \mbox{if} \quad z_0=0,\\
&\alpha= 2 \quad \mbox{if} \quad z_0=1,\\
&\alpha= -2 \quad \mbox{if} \quad z_0\neq0, 1.
\end{cases}
\label{eqn:LeadingOrder}
\end{align}
We also find that, by Eq.(\ref{eqn:FieldEqz}), $v_1,v_2,v_3,\cdots$ are determined successively in terms of  $z_0$ and $v_0$. Note that the behavior of $v(z)$ implied by Eq.(\ref{eqn:LeadingOrder}) matches the boundary condition
\begin{align}
f(0)=\pi, \quad f(\infty)=0
\label{eqn:Boundary}
\end{align}
since $z=0,1$ correspond to $r=0,\infty$, respectively. With the aid of the boundary condition (\ref{eqn:Boundary}) and the continuity of $f(r)$, we see that there should be $z_0$ such that
\begin{align} 
f(r_0)=& \frac{\pi}{2}, \quad  r_0= \sqrt{\frac{2z_0}{1-z_0}}, \quad 0<z_0<1.
\label{eqn:r0}
\end{align}
Noting that Eq.(\ref{eqn:r0}) implies $v(z_0)=\infty$, we find that $v(z)$ takes the form
\begin{align}
v(z)= \frac{z(1-z)^2}{(z-z_0)^2}w(z), \label{eqn:Profilez}
\end{align}
where $w(z)$ is a smooth function which is nonvanishing and finite for $0\leqq z\leqq 1$. Then we see that the natural relation between $f(r)$ and $v(z)$ inverse to Eq.(\ref{eqn:Vz}) is given by
\begin{align}
f(r)=\begin{cases}
&\pi-\mbox{Arctan}\sqrt{v(z)},\quad 0\leqq z \leqq z_0,\\
&\mbox{Arctan}\sqrt{v(z)},\quad z_0\leqq z \leqq 1.
\end{cases}\end{align}
Although it is difficult to obtain an exact $w(z)$, we know that it must be a moderate function for $0\leqq z\leqq 1$. Thus we may be allowed to truncate $w(z)$.\\
If we set $w(z)$ as
\begin{align}
w(z)=\sum^{N}_{k=0}w_k(z-z_0)^{k},\quad N\mbox{:small},
\end{align}
and still use the field equation for $w(z)$, then $w_1, w_2, \cdots$ are successively determined as
\begin{align}
&w_1=\frac{\left(4z_0-1\right)w_0}{2z_0\left(1-z_0\right)}, \label{eqn:w1}\\
&w_2=\frac{\left(148z_0^2-72z_0+25\right)w_0+16\left(3-z_0\right)}{48z_0^2\left(1-z_0\right)^2}, \label{eqn:w2}\\
&w_3=\frac{\left(51-200z_0+292z_0^2-408z_0^3\right)w_0+32z_0\left(3-9z_0+2z_0^2\right)}{96z_0^3\left(z_0-1\right)^3}, \label{eqn:w3}
\end{align}
and so on. In this case, $v(z)$ involves two adjustable parameters $z_0$ and $w_0$.\\
We report here the results for some small $N$. We first consider the simplest case that $w(z)$ is constant, i.e., $N=0$:
\begin{align}
w(z)=w_0.
\end{align}
The behavior of the profile function near $r=0,r_0,1$ is calculated to be
\begin{align}
f(r)& \sim\pi-\kappa_1 r, \quad  \quad \kappa_1= \frac{1}{z_0}\sqrt{\frac{w_0}{2}}, \quad(r\sim 0), \label{eqn:f1}\\
f(r)& \sim \frac{\pi}{2} -\kappa_2 (r-r_0), \quad \kappa_2= \sqrt{\frac{2(1-z_0)}{w_0}},\quad (r\sim r_0), \label{eqn:f2}\\
f(r)&\sim  \frac{\kappa_3}{r^2}, \quad \quad \quad \kappa_3= \frac{2 \sqrt{w_0}}{1-z_0},\quad (r\sim \infty) \label{eqn:f3}.
\end{align}
We have numerically estimated $E$ with changing $w_0$ and $z_0$ by the step 0.001.\\
We find that
\begin{align} 
E=1.23186\cdots\cong 1.2319 
\label{eqn:Values}
\end{align}
for
\begin{align}
v(z)=\frac{0.673z\left(1-z\right)^2}{\left(z-0.279\right)^2}
\label{eqn:Minimum}
\end{align}
is at least a local minimum. The profile function and energy density of this case is depicted in Fig.1 and Fig.2, respectively.\\

\begin{center}
\scalebox{1}[1]{\includegraphics{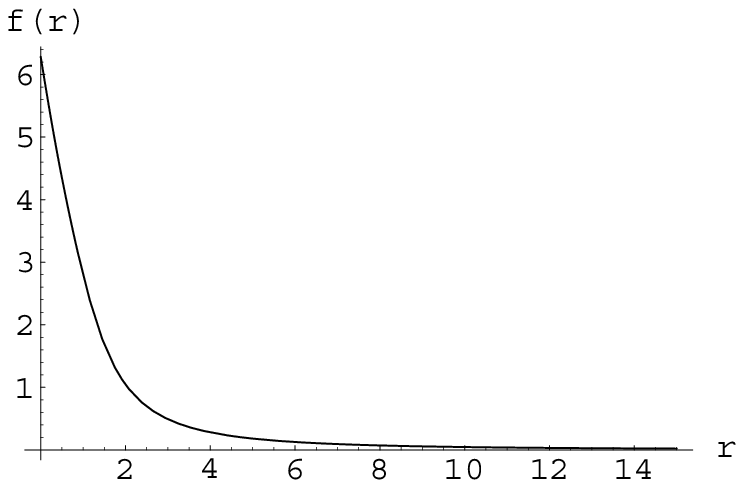}}\\
Fig.1: \begin{minipage}[t]{45mm}Profile function $f(r)$ for $w_0=0.673, z_0=0.279$.\end{minipage}\\
\scalebox{1}[1]{\includegraphics{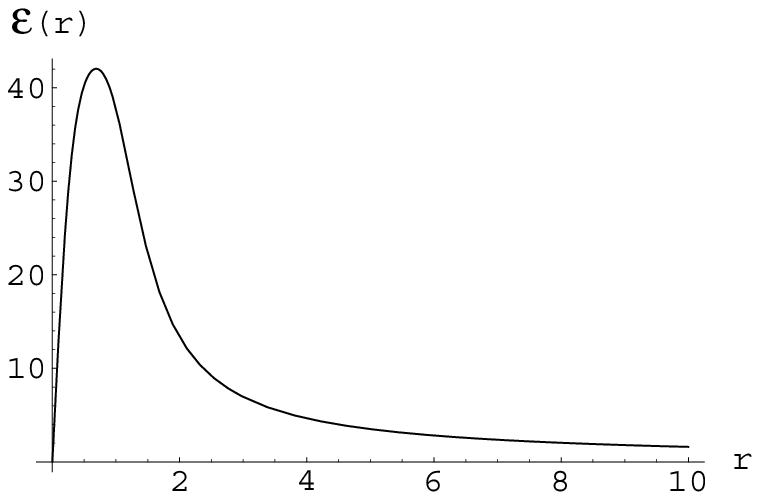}}\\
Fig.2: \begin{minipage}[t]{45mm}Energy density $\varepsilon(r)$ for $w_0=0.673, z_0=0.279$.\end{minipage}
\end{center}

In the $N=1,2,3$ cases, with the help of (\ref{eqn:w1}-\ref{eqn:w3}), we have
\begin{align}\begin{cases}
&N=1,\quad w_0=0.670,\quad  z_0=0.275,\quad E=1.23215\cdots\cong 1.2322.\\
&N=2,\quad w_0=0.485,\quad  z_0=0.327,\quad E=1.34000\cdots\cong 1.3400.\\
&N=3,\quad w_0=0.385,\quad  z_0=0.249,\quad E=1.34234\cdots\cong 1.3423.
\end{cases}\end{align}
We see that the $N=0$ case gives a smaller $E$ than $N=1,2,3$ cases.\\
On the other hand, if we do not make use of (\ref{eqn:w1}-\ref{eqn:w3}) and consider the case that $w(z)$ is a polynomial of $z$ of second order, we have the minimal energy
\begin{align}
E=1.23147\cdots\cong 1.2315
\end{align}
for
\begin{align}
v(z)=\frac{z\left(1-z\right)^2}{\left(z-0.278\right)^2}\left(0.625+0.322z-0.356z^2\right).
\end{align}
If we add some terms of the form of $z^k, k\geqq 3$ to the above $\left(0.625+0.322z-0.356z^2\right)$, we would obtain smaller $E$'s. We have thus suggested that the value of the static energy may become smaller than $1.2322$.

In ref.[2], some physical properties of nucleons were discussed in terms of the numerical solution of Eq.(\ref{eqn:FieldEq}) and the input data that masses of nucleons and delta particles are given by $M_{\mbox{N}}=939$ MeV and $M_{\Delta}=1232$ MeV, respectively. We briefly compare     here the results obtained in the above $N=0$ case ($v(z)$ of Eq.(\ref{eqn:Minimum})) with those in ref.\cite{Witten}. In ref.[2], $M_{\mbox{N}}$ and $M_{\Delta}$ are given as $M_{\mbox{N}}=M+3/(8\lambda),\quad M_{\Delta}=M+15/(8\lambda)$, where the parameters $M$ and $\lambda$ are related to $E$ of Eq.(\ref{eqn:StaticEnergy}) and $f(r)$  by
\begin{align}
&M=\frac{3{\pi}^2F_{\pi}}{e} E,\\
&\lambda=\frac{2\pi}{3e^3F_{\pi}}\Lambda,\\
&\Lambda=8\int_{0}^{\infty}dr~ r^2 \sin^2 f(r) \left[ 1+\left(\frac{df(r)}{dr}\right)^2 +\frac{\sin^2f(r)}{r^2}\right],
\end{align}
where $e$ and $F_{\pi}$ are coupling parameters appearing in ref.\cite{Witten}. The input data for $M_{\mbox{N}}$ and   $M_{\Delta}$ yield $M=866\mbox{MeV}$ and $\lambda=0.00514(\mbox{MeV})^{-1}$ up to three decimal places. The isoscalar electric mean square  radius $\sqrt{\langle r^2 \rangle_{I=0}}$ and the isoscalar magnetic mean square radius  $\sqrt{\langle r^2 \rangle_{M, I=0}}$ are defined by 
\begin{align}
&\sqrt{\langle r^2 \rangle_{I=0}}=\frac{2}{e F_{\pi}}\sqrt{-\frac{2}{\pi} \int_{0}^{\infty} dr~ r^2\sin^2f(r)f^{\prime}(r)}, \\
&\sqrt{\langle r^2 \rangle_{M, I=0}}=\frac{2}{e F_{\pi}} \sqrt{\frac{\int_{0}^{\infty} dr~ r^4\sin^2f(r)f^{\prime}(r)}{\int_{0}^{\infty} dr~ r^2\sin^2f(r)f^{\prime}(r)}}.
\end{align}
In ref.[2], the results of the analysis are stated in the following way.  
\begin{eqnarray}
M=36.5 \frac{F_{\pi}}{e},~\Lambda=50.9,~e=5.45,~F_{\pi}=129\mbox{ MeV}, \nonumber\\ 
\sqrt{\langle r^2\rangle_{I=0}}=0.59\mbox{ fm}, \quad \sqrt{\langle r^2 \rangle_{M, I=0}}=0.92\mbox{ fm},
\end{eqnarray}
while we have in the $N=0$ case 
\begin{eqnarray}
M=36.5 \frac{F_{\pi}}{e},~\Lambda=52.2,~e=5.48,~F_{\pi}=130\mbox{ MeV}, \nonumber\\ 
\sqrt{\langle r^2 \rangle_{I=0}}=0.586\mbox{ fm}, \quad \sqrt{\langle r^2 \rangle_{M, I=0}}=0.920\mbox{ fm}.
\end{eqnarray}

We find that our $N=0$ case almost reproduces the results of ref. \cite{Witten}.
The characteristic feature of our procedure is that we make use of $v(z)$ consisting of two branches of the function $\arctan\sqrt{v(z)}$. It does not cause any difficulty in the behavior of the profile function $f(r)$ as is seen in Eqs.(\ref{eqn:f1}-\ref{eqn:f3}). Because of the intimate relationship between the Faddeev and Skyrme models \cite{Battye3}, the minimal energy of the soliton of the Faddeev model \cite{Faddeev} with Hopf charge 1 may be made smaller.
\section*{Acknowledgments}
The authors are grateful to Kouichi Toda for discussions. The authors also thank Nobuyuki Sawado, Hitoshi Yamakoshi, Shinji Hamamoto, Takeshi Kurimoto, and Hiroshi Kakuhata for discussions. This work was supported in part by a Japanese Grant-in-Aid for Scientific Research from the Ministry of Education, Culture, Sports, Science and Technology (No.13135211).

\end{document}